\documentclass[10pt]{article}
\usepackage{latexsym,graphicx, amssymb}
\newcommand{\be}{\begin{equation}}
\newcommand{\ee}{\end{equation}}
\def\n{\noindent}
\catcode `\@=11
\catcode `\@=12
\begin{document}
\begin{center}
\large{\bf{String Cosmology in LRS Bianchi Type-II Dusty Universe with Time Decaying Vacuum Energy Density $\Lambda$}} \\
\vspace{10mm}
\normalsize{Hassan Amirhashchi}\\
\vspace{5mm} \normalsize{Department of Physics, Islamic Azad University, Mahshahr Branch, Mahshahr, Iran\\
E-mail : h.amirhashchi@mahshahriau.ac.ir}, hashchi@yahoo.com \\

\end{center}
\vspace{10mm}
\begin{abstract}
A model of a cloud formed by massive strings is used as a source
of LRS Bianchi type II with time decaying vacuum energy density $\Lambda$. To construct string cosmological models we have used
the energy-momentum tensor for such string as formulated by Letelier (1983). The high nonlinear field equations have been solved for two
types of strings, (i) Massive string and (ii) Nambu string. The expansion $\theta$ in the model is assumed to be proportional to the shear
$\sigma$. This condition leads to $A=\beta B^{m}$, where $A$ and $B$ are the metric coefficients, $m$ is a constant and $\beta$ is an integrating
constant. Our models are in accelerating phase which is consistent to the recent observations of supernovae type Ia. The physical and geometrical
behavior of these models are also discussed.
\end{abstract}
\smallskip
\n Keywords: LRS Bianchi type II models. Nambu String. Massive string\\
\n PACS number: 98.80.Cq, 04.20.-q, 04.20.Jb, 98.80.-k
\section{Introduction}
The Problem of the structure formation in the universe is an important challenge in cosmology. There are two theories for the structure formation of the Universe, (i) first one is based upon the amplification of quantum fluctuations in a scalar field during inflation and (ii) second one is based upon symmetry
breaking phase transition in the early Universe which leads to the formation of {\it{topological defects}} such as domain walls, cosmic
strings, monopoles, textures and other ``hybrid" creatures. However, domain walls and monopoles are disastrous for the cosmological models but Strings, on the other hand, causes no harm, but can lead to very interesting astrophysical consequences \cite{ref1}.
The cosmic strings play a vital role in the formation of galaxies. They have been created when the symmetry between the strong and electroweak forces broken due to the phase transition in the early universe ($t\backsim 10^{-36}s$) \cite{ref1} as the temperature falls down below some critical temperature ($T_{GUT}=10^{28}K$) as predicted by grand unified theories (GUT) \cite{ref1}-\cite{ref6}. The existence
of a large scale network of strings the early universe does not contradict the present-day observations. The vacuum strings
may generate density fluctuations sufficient to explain the galaxy formation \cite{ref7}. The cosmic strings have coupled stress-energy to the gravitational field. Therefor, the study of gravitational effects of such strings will be interesting. The general relativistic treatment of strings was initiated by Letelier \cite{ref8,ref9}. Here we have considered gravitational effects, arisen from strings by coupling
of stress energy of strings to the gravitational field. Letelier (1983) \cite{ref9} defined the massive strings as the geometric strings
(massless) with particles attached along its expansions.\\
The strings that form the cloud are massive strings instead of geometrical strings. Each massive string is formed
by a geometrical string with particles attached along its extension. Hence, the string that form the cloud are the generalization
of Takabayasi's relativistic model of strings (called $p$-strings). This is simplest model wherein we have particles
and strings together. In principle we can eliminate the strings and end up with a cloud of particles. This a desirable property
of a model of a string cloud to be used in cosmology since strings are not observed at the present time of evolution
of the universe (see \cite{ref10}-\cite{ref13}).\\

There are significant observational evidence that the expansion of the Universe is undergoing a late time acceleration (Perlmutter {\it et
al.} \cite{ref14}$-$\cite{ref15}; Riess {\it et al.} \cite{ref16,ref17}; Garnavich {\it et al.} \cite{ref18,ref19}; Schmidt {\it et al.}
\cite{ref20}; Efstathiou {\it et al.} \cite{ref21}; Spergel {\it et al.} \cite{ref22}; Allen et al. \cite{ref23}; Sahni and Starobinsky
\cite{ref24}; Peebles and Ratra \cite{ref25}; Padmanabhan \cite{ref26}; Lima \cite{ref27}). This, in other words, amounts to
saying that in the context of Einstein's general theory of relativity some sort of dark energy, constant or that varies only
slowly with time and space dominates the current composition of cosmos. The origin and nature of such an accelerating field poses a
completely open question. Recently Riess {\it et al.} \cite{ref28} have presented an analysis of 156 SNe including a few at $z > 1.3$
from the Hubble Space Telescope (HST) ``GOOD ACS'' Treasury survey. They conclude to the evidence for present acceleration $q_{0} < 0$
$(q_{0} \approx -0.7)$. Observations (Riess {\it et al.}; \cite{ref30}; Knop et al. \cite{ref29}) of Type Ia Supernovae (SNe) allow
us to probe the expansion history of the universe leading to the conclusion that the expansion of the universe is accelerating.
Observations strongly favor a small and positive value of the effective cosmological constant with magnitude
 $\Lambda(G\hbar/c^{3})\approx 10^{-123}$ at the present epoch.\\
Some of the recent discussions on the cosmological constant ``problem'' and on cosmology
with a time-varying cosmological constant point out that in the absence of any interaction with matter or radiation,
the cosmological constant remains a "constant". However, in the presence of interactions with matter or radiation, a
solution of Einstein equations and the assumed equation of covariant conservation of stress-energy with a time-varying $\Lambda$ can be found.
This entails that energy has to be conserved by a decrease in the energy density of the vacuum component
followed by a corresponding increase in the energy density of matter or radiation.\\

In recent times the study of anisotropic string cosmological models have generated a lot of research interest.
Reddy \cite{ref31,ref32}, Reddy and Naidu \cite{ref33}, Reddy et al. \cite{ref34,ref35}, Rao et al. \cite{ref36}-\cite{ref39},
Pradhan \cite{ref40,ref41}, Pradhan and Mathur \cite{ref42},
Pradhan et al. \cite{ref43}-\cite{ref46}, Amirhashchi and Hishamuddin
\cite{ref47}-\cite{ref49}, Tripathi et al. \cite{ref50,ref51} have studied
string cosmological models in different contexts. Recently,
Amirhashchi and Zainuddin \cite{ref47} have obtained LRS Bianchi type
II strings dust cosmological models for perfect fluids distribution in general relativity.
Motivated the situations discussed above, in this paper, we shall focus upon the problem of
establishing a formalism for studying the massive string in LRS Bianchi II space-time. In this paper, we have investigated a new and
general solution for Bianchi type-III cosmological model for a cloud of strings and time varying cosmological constan which is different from the
other solutions. The paper is organized as follows. The metric and
the field equations are presented in Section 2. In Section 3, we
deal with solution of the field equations with cloud of strings and
cosmological constant. In Subsec 3.1 the field equations have been
solved for case of massive string. We describe some physical and
geometric properties of the model in Subsec 3.1.1. In Subsec 3.2 the field equations have been solved for case
of Nambu or geometric string. We describe some physical and
geometric properties of the model in Subsec 3.2.1. Finally, in Section 4, concluding remarks are given.
\section{The Metric and Field Equations}
We consider the LRS Bianchi type II metric in the form
\begin{equation}
\label{eq1} ds^{2}=-dt^{2}+B^{2}(dx+zdy)^{2}+A^{2}(dy^{2}+dz^{2}),
\end{equation}
where $A$ , $B$ are functions of $t$ only. The energy momentum tensor for a cloud of strings is taken as
\begin{equation}
\label{eq2} T_{ij}=\rho u_{i}u_{j}-\lambda x_{i}x_{j},
\end{equation}
where $u_{i}$ and $x_{i}$ satisfy condition
\begin{equation}
\label{eq3} u_{i}u^{i}=-x^{i}x_{i}=-1,\quad u^{i}x_{i}=0,
\end{equation}
where $\rho$ is the proper energy density for a cloud
string with particles attached to them, $\lambda$ is the string tension density, $u^{i}$ the
four-velocity of the particles, and $x^{i}$ is a unit space-like vector representing
the direction of string. In a co-moving coordinate system, we have
\begin{equation}
\label{eq4} u^{i}=(0,0,0,1),\quad x^{i}=(\frac{1}{B},0,0,0).
\end{equation}
The particle density of the configuration is given by
\begin{equation}
\label{eq5} \rho=\rho_{p}+\lambda.
\end{equation}
The Einstein's field equations (with $8\pi G=1$ and $C=1$)
\begin{equation}
\label{eq6} R_{ij}-\frac{1}{2}Rg_{ij}+\Lambda g_{ij}=-T_{ij},
\end{equation}
for the metric (\ref{eq1}) leads to the following system of equations:
\begin{equation}
\label{eq7} 2\frac{\ddot{A}}{A}+\frac{\dot{A}^{2}}{A^{2}}-\frac{3}{4}\frac{B^{2}}{A^{4}}=\lambda-\Lambda;
\end{equation}
\begin{equation}
\label{eq8} \frac{\ddot{A}}{A}+\frac{\ddot{B}}{B}+\frac{\dot{A}\dot{B}}{AB}+\frac{1}{4}\frac{B^{2}}{A^{4}}=-\Lambda;
\end{equation}
\begin{equation}
\label{eq9}2\frac{\dot{A}\dot{B}}{AB}+\frac{\dot{A}^{2}}{A^{2}}-\frac{1}{4}\frac{B^{2}}{A^{4}}=\rho-\Lambda,
\end{equation}
where an overdot stands for the first and double overdot for second derivative with respect to $t$.\\
The average scalar factor $R$ for LRS Bianchi type II is given by
\begin{equation}
\label{eq10}a=(A^{2}B)^{\frac{1}{3}}.
\end{equation}
A volume scale factor $V$ is defined as
\begin{equation}
\label{eq11}V=a^{3}=A^{2}B.
\end{equation}
We define the generalized mean Hubble's parameter $H$ as
\begin{equation}
\label{eq12}H=\frac{1}{3}(H_{1}+H_{2}+H_{3}),
\end{equation}
where $H_{1}=H_{2}=\frac{\dot{A}}{A}$ and $H_{3}=\frac{\dot{B}}{B}$ are the directional Hubble's parameters in the directions of $x, y$ and $z$ respectively.\\
From (\ref{eq11}) and (\ref{eq12}), we obtain
\begin{equation}
\label{eq13}H=\frac{1}{3}\frac{\dot{V}}{V}=\frac{\dot{a}}{a}=\frac{1}{3}\left(\frac{\dot{B}}{B}+2\frac{\dot{A}}{A}\right).
\end{equation}
An important observational quantity is the deceleration parameter $q$, which is defined as
\begin{equation}
\label{eq14}q=-\frac{\ddot{a}a}{\dot{a}^{2}}.
\end{equation}\\
The scalar expansion $\theta$, the shear scalar $\sigma^{2}$ and the average anisotropy parameter $A_{m}$ are defined as
\begin{equation}
\label{eq15}\theta=u^{i}_{;i}= \frac{\dot{B}}{B}+2\frac{\dot{A}}{A},
\end{equation}
\begin{equation}
\label{eq16} \sigma^{2}=\frac{1}{2}\sigma_{ij}\sigma^{ij}=\frac{1}{2}\left[2\frac{\dot{A}}{A}+\frac{\dot{B}^{2}}{B^{2}}\right]-\frac{\theta^{2}}{6},
\end{equation}
\begin{equation}
\label{17} A_{m}=\frac{1}{3}\sum^{3}_{i}\left(\frac{\triangle H_{i}}{H}\right)^{2},
\end{equation}
where $\triangle H_{i}=H_{i}-H (i=1, 2, 3)$.
\section{Solution of the Field Equations}
The field equations (\ref{eq7})-(\ref{eq9}) are a system of three equations with five unknown parameters $A, B, \rho, \lambda, \Lambda$. Two additional constraints relating these parameters are required to obtain explicit solutions of the system. If we assume that the expansion $\theta$ in the model is proportional to the shear $\sigma$. This condition leads to
\begin{equation}
\label{eq18} A=\beta B^{m},
\end{equation}
where  $m$ is a constant and $\beta$ is a constant of integration.\\
From Eqs. (\ref{eq7}) and (\ref{eq9}) we get
\begin{equation}
\label{eq19} \lambda=2\frac{\ddot{A}}{A}+\frac{\dot{A}^{2}}{A^{2}}-\frac{3}{4}\frac{B^{2}}{A^{4}}+\Lambda,
\end{equation}
and
\begin{equation}
\label{eq20} \rho=2\frac{\dot{A}\dot{B}}{AB}+\frac{\dot{A}^{2}}{A^{2}}-\frac{1}{4}\frac{B^{2}}{A^{4}}+\Lambda.
\end{equation}
The highly non-linear field equations can be solved for the following two types of
geometric strings.
\subsection{ Case I: Massive String}
In this case we assume that the sum of rest energy density and tension density for cloud of strings vanish \cite{ref52}-\cite{ref54}. i.e.
\begin{equation}
\label{eq21} \rho+\lambda=0.
\end{equation}
From Eqs. (\ref{eq19}), (\ref{eq20}) and (\ref{eq21}) we obtain
\begin{equation}
\label{eq22} 2\frac{\dot{A}\ddot{B}}{AB}+2\frac{\dot{A}^{2}}{A^{2}}+2\frac{\ddot{A}}{A}=\frac{B^{2}}{A^{4}}-2\Lambda.
\end{equation}
From (\ref{eq18}) equation (\ref{eq22}) can be written as
\begin{equation}
\label{eq23} 2\ddot{B}+4m\frac{\dot{B}^{2}}{B}=\frac{B^{3-4m}}{m\beta}-\frac{2\Lambda}{m}B.
\end{equation}
Let $\dot{B}=f(B)$ which implies that $\ddot{B}=ff'$, where $f'=\frac{df}{dB}$. Hence (\ref{eq23}) takes the form
\begin{equation}
\label{eq24} \frac{d}{dB}(f^{2})+\frac{4m}{B}f^{2}=\frac{B^{3-4m}}{m\beta}-\frac{2\Lambda}{m}B.
\end{equation}
Eq. (\ref{eq24}), after integrating, reduces to
\begin{equation}
\label{eq25} f^{2}=(\frac{dB}{dt})^{2}=\frac{1}{4\beta m}B^{4(1-m)}-\frac{\Lambda}{m(1+2m)}B^{2}+NB^{-4m},
\end{equation}
where $N$ is an integrating constant. The equation (\ref{eq25}) can be written in the following form
\begin{equation}
\label{eq26} f^{2}=aB^{4(1-m)}-bB^{2}+NB^{-4m}.
\end{equation}
To get deterministic solution, we assume $m=\frac{1}{2} $. In this case (\ref{eq26}) takes the form
\begin{equation}
\label{eq27} f^{2}=MB^{2}+NB^{-2},
\end{equation}
where
\begin{equation}
\label{eq28} M=a-b.
\end{equation}
Therefore, we find
\begin{equation}
\label{eq29} \frac{dB}{\sqrt{MB^{2}+NB^{-2}}}=dt.
\end{equation}
Integrating (\ref{eq29}), we obtain
\begin{equation}
\label{eq30} B^{2}=(\frac{N}{M})^{\frac{1}{2}}sinh[2\sqrt{M}(t+\alpha)],
\end{equation}
and
\begin{equation}
\label{eq31} A^{2}=\beta^{2}(\frac{N}{M})^{\frac{1}{4}}sinh^{\frac{1}{2}}[2\sqrt{M}(t+\alpha)],
\end{equation}
where $M> 0$ without any loss of generality.\\
Thus the metric (\ref{eq1}) reduces to
\[
ds^{2}=-dt^{2}+(\frac{N}{M})^{\frac{1}{2}}sinh[2\sqrt{M}(t+\alpha)](dx+zdy)^{2}
\]
\begin{equation}
\label{eq32} +\ \beta^{2}(\frac{N}{M})^{\frac{1}{4}}sinh^{\frac{1}{2}}[2\sqrt{M}(t+\alpha)](dy^{2}+dz^{2}).
\end{equation}
After using a suitable transformation of coordinates the model (\ref{eq32}) reduces to
\[
ds^{2}=-dt^{2}+(\frac{N}{M})^{\frac{1}{2}}sinh(2\sqrt{M}T)(dx+zdy)^{2}
\]
\begin{equation}
\label{eq33}+\ \beta^{2}(\frac{N}{M})^{\frac{1}{4}}sinh^{\frac{1}{2}}(2\sqrt{M}T)(dy^{2}+dz^{2}).
\end{equation}
\subsection*{3.1.1 The Geometric and Physical Significance of Model}
The energy density $(\rho)$, the string tension $(\lambda)$, the particle density $(\rho_{p})$ and the vacuum energy
 density $(\Lambda)$ for the model (\ref{eq33}) are given by
\begin{figure}[htbp]
\centering
\includegraphics[width=8cm,height=8cm,angle=0]{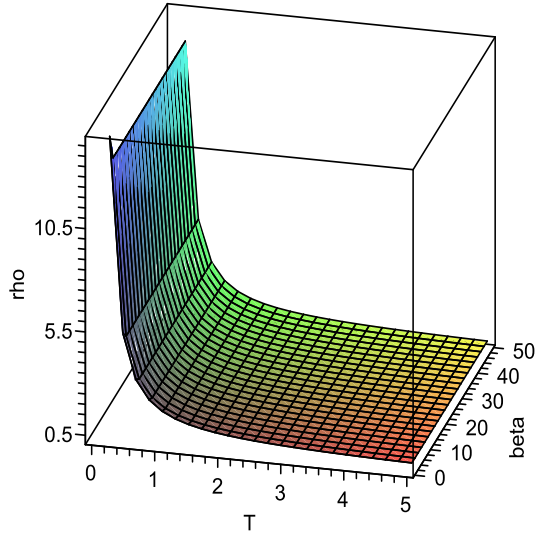}
\caption{The plot of energy density $\rho$ Vs. T and $\beta$}
\end{figure}
\begin{equation}
\label{eq34} \rho=\frac{5}{4}Mcoth^{2}(2\sqrt{M}T)-\frac{1}{4\beta^{2}}+\Lambda
\end{equation}
\begin{figure}[htbp]
\centering
\includegraphics[width=8cm,height=8cm,angle=0]{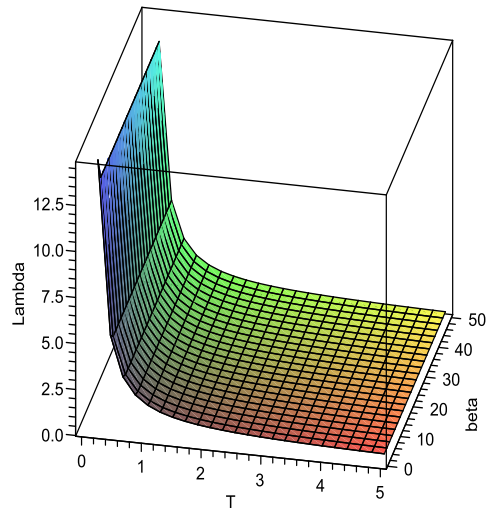}
\caption{The plot of vacuum energy density $\Lambda$ Vs. T and $\beta$}
\end{figure}
\begin{equation}
\label{eq35} \lambda=-\frac{5}{4}Mcoth^{2}(2\sqrt{M}T)+\frac{1}{4\beta^{2}}-\Lambda
\end{equation}
\begin{figure}[htbp]
\centering
\includegraphics[width=8cm,height=8cm,angle=0]{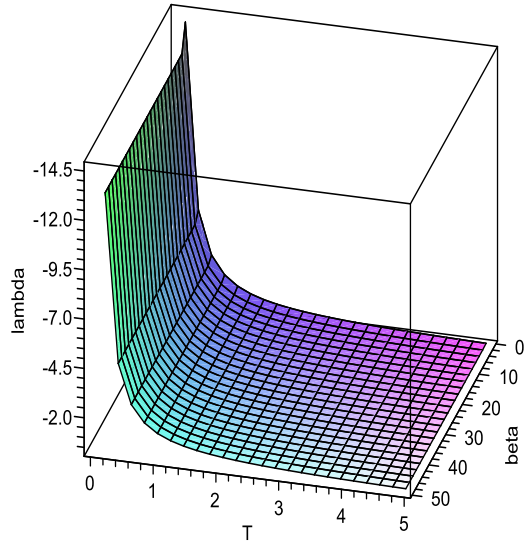}
\caption{The plot of tension density $\lambda$ Vs. T and $\beta$}
\end{figure}
\begin{equation}
\label{eq36} \rho_{p}=\frac{5}{2}Mcoth^{2}(2\sqrt{M}T)-\frac{1}{2\beta^{2}}+ 2\Lambda
\end{equation}
\begin{equation}
\label{eq37} \Lambda= \frac{\left( 8\, \left( -1/4\,{\beta}^{-2}-3/2\,{\beta}^{-1} \right)
\beta+5\, \left( \coth \left( 2\sqrt{M} T \right)  \right) ^{2} \right)}{\left(
-16\,\beta+10\, \left( \coth \left( 2\sqrt{M} T \right)  \right) ^{2}\beta
 \right)}.
\end{equation}
From (\ref{eq34}) and (\ref{eq36}), we see that energy conditions, $\rho\geq 0$ and $\rho_{p}\geq 0$ are satisfied under conditions
\begin{equation}
\label{eq38} coth^{2}(2\sqrt{M}T)\geq\frac{1-4\beta^{4}\Lambda}{5M\beta^{4}}.
\end{equation}
We also observe that string tension density $\lambda\geq 0$ and cosmological constant
$\Lambda\geq 0$ under conditions
\begin{equation}
\label{eq39} coth^{2}(2\sqrt{M}T)\leq\frac{1-4\beta^{4}\Lambda}{5M\beta^{4}}.
\end{equation}
and
\begin{equation}
\label{eq40} \frac{coth^{2}(2\sqrt{M}T)}{2}\geq(\frac{1+3\beta}{\beta})\geq\frac{8}{10} ~ ~or ~ ~ \frac{coth^{2}(2\sqrt{M}T)}{2}\leq(\frac{1+3\beta}{\beta})\leq\frac{8}{10},
\end{equation}
respectively. From Eq. (\ref{eq34}), it is noted that the proper energy density $\rho(t)$ is a decreasing
function of time and it approaches a small positive value at present epoch. This behavior is clearly depicted in Figures 1 as a
representative case with appropriate choice of constants of integration and other physical parameters using
reasonably well known situations.\\

From Eq. (\ref{eq35}) and also by comparing Eqs. (\ref{eq38}) and (\ref{eq39}), it is found that the tension density $\lambda$ is negative. From Figure
3, it is observed that $\lambda$ is a decreasing function of time and it is always negative.
It is pointed out by Letelier \cite{ref9} that $\lambda$ may be positive or negative. When $\lambda < 0$,
the string phase of the universe disappears i.e. we have an anisotropic fluid of particles.\\

 From Eq. (\ref{eq37}), we see that the cosmological term $\Lambda$ is a decreasing function of time and it approaches
a small positive value at late time. From Figure 2, we note this behavior of cosmological term $\Lambda$
in the model. Recent cosmological observations  suggest the existence of a positive cosmological constant
$\Lambda$ with the magnitude $\Lambda(G\hbar/c^{3})\approx 10^{-123}$. These observations on magnitude and
red-shift of type Ia supernova suggest that our universe may be an accelerating one with induced cosmological
density through the cosmological $\Lambda$-term. Thus, our model is consistent with the results of recent
observations. \\

The expressions for the scalar of expansion $\theta$, magnitude of shear $\sigma^{2}$, proper volume $V$, deceleration parameter $q$ and
the average anisotropy parameter $A_{m}$ for the model (\ref{eq33}) are given by
\begin{equation}
\label{eq41} \theta^{2}=4Mcoth^{2}(2\sqrt{M}T),
\end{equation}
\begin{equation}
\label{eq42} \sigma^{2}=\frac{1}{12}Mcoth^{2}(2\sqrt{M}T),
\end{equation}
\begin{equation}
\label{eq43} V=\beta^{2}(\frac{N}{M})^{\frac{1}{2}}sinh(2\sqrt{M}T),
\end{equation}
\begin{equation}
\label{eq44} q=-\left[\frac{\frac{4M}{3}-\frac{8M}{9}coth^{2}(2\sqrt{M}T)}{\frac{4}{9}Mcoth^{2}(2\sqrt{M}T)}\right],
\end{equation}
\begin{equation}
\label{eq45}A_{m}=2(\frac{m-1}{2m+1})^{2}=\frac{1}{8}.
\end{equation}
The rate of expansion $H_{i}$ in the direction of $x$, $y$ and $z$ are given by
\begin{equation}
\label{eq46}H_{1}=H_{2}=\frac{\sqrt{M}}{2}coth(2\sqrt{M}T), ~ ~ H_{3}=\sqrt{M}coth(2\sqrt{M}T).
\end{equation}
Hence the average generalized Hubble's parameter is given by
\begin{equation}
\label{eq47} H=(\frac{2m+1}{3})\sqrt{M}coth(2\sqrt{M}T)=\frac{2}{3}\sqrt{M}coth(2\sqrt{M}T).
\end{equation}
To indicate whether a model inflates or not one can finds sign of $q$. A negative sign $-1 \leq q<0$, indicates inflations whereas positive sign of $q$ corresponds to standard decelerating model.
From (35), we observe that
\begin{equation}
\label{eq48} q<0 \qquad if\quad coth^{2}(2\sqrt{M}T)<\frac{3}{2},
\end{equation}
and
\begin{equation}
\label{eq49} q>0 \qquad if\quad coth^{2}(2\sqrt{M}T)>\frac{3}{2}.
\end{equation}

The model (\ref{eq33}) starts with a big bang at $T=0$. The expansion in the model decreases as time increases. The proper volume of the model increases as time increases. Since $\frac{\sigma}{\theta}$ is constant the model does not approach isotropy. There is a point type singularity in the model at $T=0$ (MacCallum \cite{ref55}). For the condition $coth^{2}(2\sqrt{M}T)<\frac{3}{2}$, the solution gives accelerating model of the universe and for the condition $coth^{2}(2\sqrt{M}T)>\frac{3}{2}$,
our solution represents decelerating model of the universe. It is also observed that at
$T=\frac{1}{2\sqrt{M}}coth^{-1}(\sqrt{3})$, $q$ approaches the value $-1$
as in the case of de-Sitter universe.\\

In this case from Eqs. (\ref{eq35}) and (\ref{eq36}), we obtain
\begin{equation}
\label{eq50} \frac{\rho_{p}}{|\lambda|}=2>1.
\end{equation}
Thus, in our model, the universe is dominated by massive strings throughout the
whole process of evolution \cite{ref1,ref56}.

\subsection{Case II: Nambu String}
In this case we assume
\begin{equation}
\label{eq51} \rho-\lambda=0.
\end{equation}
This corresponds to the state equation for a cloud of massless geometric (Nambu) strings i.e.$\rho_{p}=0$.
Therefor in this case from Eqs. (\ref{eq19}) and (\ref{eq20}), we obtain
\begin{equation}
\label{eq52} 2\frac{\dot{A}\dot{B}}{AB}-2\frac{\ddot{A}}{A}+\frac{1}{2}\frac{B^{2}}{A^{4}}=0.
\end{equation}
Putting (\ref{eq18}) in (\ref{eq52}) we get
\begin{equation}
\label{eq53} 2\ddot{B}+2(m-2)\frac{\dot{B}^{2}}{B}=\frac{B^{3-4m}}{2m\beta^{4}}.
\end{equation}
Let $\dot{B}=f(B)$ which implies that $\ddot{B}=ff'$, where $f'=\frac{df}{dB}$. This leads to
\begin{equation}
\label{eq54} \frac{d}{dB}(f^{2})+\frac{2(m-2)}{B}f^{2}=\frac{B^{3-4m}}{2m\beta^{4}}.
\end{equation}
Eq. (\ref{eq54}), after integrating, reduces to
\begin{equation}
\label{eq55} \frac{dB}{dt}=\sqrt{-\frac{B^{4(1-m)}}{4m^{2}\beta^{4}}+LB^{-2(m-2)}},
\end{equation}
where $L$ is an integrating constant.\\
Hence the metric (\ref{eq1}) reduces to
\begin{equation}
\label{eq56} ds^{2}=\frac{dB^{2}}{\left(-\frac{B^{4(1-m)}}{4m^{2}\beta^{4}}+LB^{-2(m-2)}\right)}+B^{2}(dx+zdy)^{2}+\beta^{2}B^{2m}(dy^{2}+dz^{2}).
\end{equation}
After making suitable transformation of coordinates the metric (\ref{eq56}) reduces to
\begin{equation}
\label{eq57} ds^{2}=\frac{dT^{2}}{\left(-\frac{T^{4(1-m)}}{4m^{2}\beta^{4}}+LT^{-2(m-2)}\right)}+T^{2}(dx+zdy)^{2}+\beta^{2}T^{2m}(dy^{2}+dz^{2}).
\end{equation}
\subsection*{3.2.1 The Geometric and Physical Significance of Model}

The energy density $(\rho)$, the string tension $(\lambda)$, the particle density $(\rho_{p})$, the vacuum energy density $\Lambda$, the scalar of expansion $(\theta)$, the shear $(\sigma)$ and the proper volume $(V^{3})$ for the model (\ref{eq57}) are given by
\[
\rho=\lambda=m(m+2)\left(\frac{1}{4m^{2}\beta^{4}}-LT^{2m}\right)^{2}T^{-2(4m-3)}
\]
\begin{equation}
\label{eq58} \ +\frac{1}{4\beta^{4}}T^{-2(2m-1)}+\Lambda,
\end{equation}
\begin{figure}[htbp]
\centering
\includegraphics[width=8cm,height=8cm,angle=0]{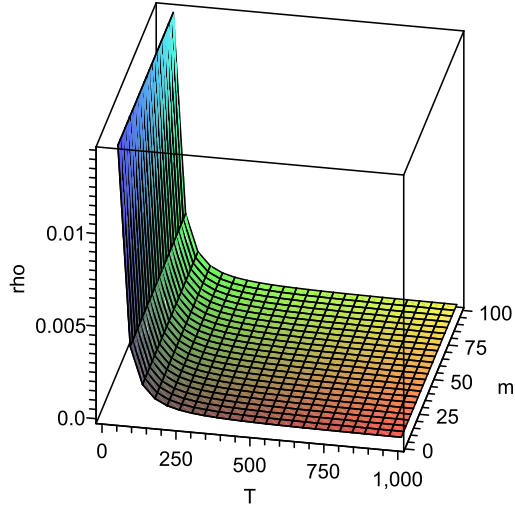}
\caption{The plot of energy density $\rho$ Vs. T and $m$}
\end{figure}
\begin{figure}[htbp]
\centering
\includegraphics[width=8cm,height=8cm,angle=0]{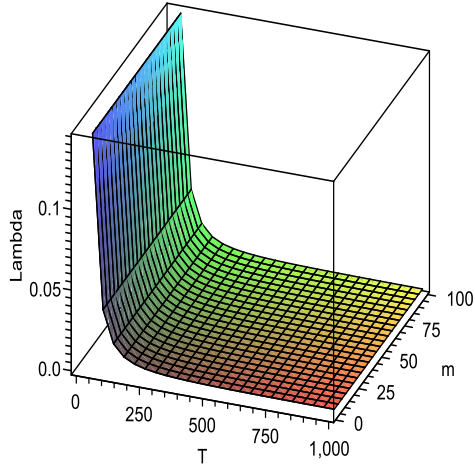}
\caption{The plot of cosmological constant $\Lambda$ Vs. T and $m$}
\end{figure}
\begin{equation}
\label{eq59}\rho_{p}=0,
\end{equation}
\[
\Lambda=\left[(\frac{3m^{2}-4}{16m^{4}\beta^{8}})-(\frac{2m^{2}+m+4}{2m^{2}\beta^{4}})LT^{2m}-(m^{2}-2m+4)L^{2}T^{4m}\right]T^{-2(4m-3)}
\]
\begin{equation}
\label{eq60}\ - \frac{1}{4\beta^{4}}T^{-2(2m-1)}.
\end{equation}
From (\ref{eq58}), we see that the energy condition, $\rho> 0$ is satisfied under condition
\begin{equation}
\label{eq61}m(m+2)>0.
\end{equation}
From Eq. (\ref{eq60}), it is found that the vacuum energy density
$\Lambda > 0$ when $T \geq T_{c}$, where $T_{c}$ is a critical time
given by
\begin{equation}
\label{eq62}T_{c}=\left[\frac{1}{2\beta^{2}\sqrt{\left[(\frac{3m^{2}-4}{16m^{4}\beta^{8}})
-(\frac{2m^{2}+m+4}{2m^{2}\beta^{4}})LT^{2m}-(m^{2}-2m+4)L^{2}T^{4m}\right]}}\right]^{\frac{1}{2(m-1)}}.
\end{equation}
From Eqs. (\ref{eq58}) and (\ref{eq60}), it is noted that the proper energy density $\rho(t)$ and the vacuum energy density $\Lambda(t)$ are decreasing
functions of time and they approach a small positive value at present epoch. This behavior is clearly depicted in Figures 4 and 5.\\
From equation (\ref{eq60}), it is clear that cosmological constant $\Lambda$ is decreasing function
of time and approaches to small positive value at late time which is supported
by results from supernova observations recently obtained by High-z Supernova
Team and Supernova cosmological project \cite{ref14}-\cite{ref21}.\\

The expressions for the scalar of expansion $\theta$, magnitude of shear $\sigma^{2}$, proper volume $V$, deceleration parameter $q$ and
the average anisotropy parameter $A_{m}$ for the model (\ref{eq57}) are given by
\begin{equation}
\label{eq63} \theta=(2m+1)\left(-\frac{1}{4m^{2}\beta^{4}}+LT^{2m}\right)T^{3-4m},
\end{equation}
\begin{equation}
\label{eq64} \sigma^{2}=\frac{(2m-1)^{2}}{\sqrt{3}}\left(-\frac{1}{4m^{2}\beta^{4}}+LT^{2m}\right)^{2}T^{-2(4m-3)},
\end{equation}
\begin{equation}
\label{eq65} V=\beta^{2}T^{2m+1},
\end{equation}
\begin{equation}
\label{eq66} q=-\frac{3}{2m+1}\left[\frac{(\frac{1-m}{48m^{4}\beta^{8}})+(\frac{7m-10}{6m^{2}\beta^{4}})LT^{2m}
+(\frac{10-4m}{3})L^{2}T^{4m}}{\left(-\frac{1}{4m^{2}\beta^{4}}+LT^{2m}\right)^{2}}\right],
\end{equation}
\begin{equation}
\label{eq67}A_{m}=2(\frac{m-1}{2m+1})^{2}.
\end{equation}
The rate of expansion $H_{i}$ in the direction of $x$, $y$ and $z$ are given by
\[
H_{1}=H_{2}=m\left(-\frac{1}{4m^{2}\beta^{4}}+LT^{2m}\right)T^{3-4m},
\]
\begin{equation}
\label{eq68}\ H_{3}=\left(-\frac{1}{4m^{2}\beta^{4}}+LT^{2m}\right)T^{3-4m}.
\end{equation}
Hence the average generalized Hubble's parameter is given by
\begin{equation}
\label{eq69} H=(\frac{2m+1}{3})\left(-\frac{1}{4m^{2}\beta^{4}}+LT^{2m}\right)T^{3-4m}.
\end{equation}
From eq. (\ref{eq66}), we observe that
\begin{equation}
\label{eq70}q<0 ~ ~ ~ if ~ ~ T^{4m}\left[2(2m-5)L-(\frac{7m-10}{2m^{2}\beta^{4}})T^{-2m}\right]<\left(\frac{1-m}{16m^{4}L\beta^{8}}\right),
\end{equation}
and
\begin{equation}
\label{eq71}q>0 ~ ~ ~ if ~ ~ T^{4m}\left[2(2m-5)L-(\frac{7m-10}{2m^{2}\beta^{4}})T^{-2m}\right]>\left(\frac{1-m}{16m^{4}L\beta^{8}}\right).
\end{equation}
Also we note that for
\begin{equation}
\label{eq72} T^{4m}\left[(2m-3)L-3(\frac{m-1}{2m^{2}\beta^{4}})T^{-2m}\right]=\left(\frac{m+2}{48m^{4}L\beta^{8}}\right),
\end{equation}
$q=-1$ as in the case of de-Sitter universe.\\

The model (\ref{eq57}) start with a big bang at $T=0$. For $m>\frac{3}{2}$, the expansion in the model decrease as time increases.
The proper volume of the model increases as time increases. There is a point type singularity in the model at T = 0 \cite{ref55}.
Since $\frac{\sigma}{\theta}$ is constant this model does not approach isotropy. From (\ref{eq66}) it is clear that our model
represents an accelerating universe for the condition given by Eq. (\ref{eq70}) and a decelerating model of the universe under condition
given by Eq. (\ref{eq71}).\\

In this case from Eqs. (\ref{eq58}) and (\ref{eq59}), we obtain
\begin{equation}
\label{eq68} \frac{\rho_{p}}{|\lambda|}=0
\end{equation}
Hence, in this case the strings dominate over the particles.

\section{Conclusion}
In this paper we have presented a new solution of Einstein's field
equations for LRS Bianchi type-II space-time with a cloud of strings in
presence of time varying cosmological constant. We have
considered two cases (i) Massive String and (ii) Nambu String. In both cases our models
starts with a big bang at $T = 0$ In both cases the models have point type big bang
singularity at $T = 0$. In both cases our models are in accelerating phase under appropriate
conditions. In case (i) the universe is dominated by massive strings
throughout the whole process of evolution. But it is observed that in case (ii) the string
dominates over the particle. In both cases the models do not
approach isotropy. Our models are realistic and new to the others.\\
It is also possible to describe cosmological constant on the basis of thermodynamics. By thermodynamics we know that
\begin{equation}
\label{eq69} \tau dS=d(\rho V^{3})+\rho dV^{3},
\end{equation}
where $V^{3}$ is proper volume and $\tau$ is temperature. Therefore
\begin{equation}
\label{eq70} \tau dS=\dot{\rho}+\rho\left(\frac{\dot{B}}{B}+2\frac{\dot{A}}{A}\right)-\lambda\frac{\dot{A}}{A}.
\end{equation}
Since in Riemannian geometry without cosmological constant we have
\begin{equation}
\label{eq71} \tau dS=0,
\end{equation}
therefor from (\ref{eq70}) we find
\begin{equation}
\label{eq72}\dot{\rho}+\rho\left(\frac{\dot{B}}{B}+2\frac{\dot{A}}{A}\right)-\lambda\frac{\dot{A}}{A}=0
\end{equation}
Since $\dot{\rho}>0$, hence from (\ref{eq72}) we conclude
\begin{equation}
\label{eq73} 3\rho H<\lambda\frac{\dot{A}}{A}.
\end{equation}
Therefor, since in both cases $\lambda$ can be negative, we conclude
\begin{equation}
\label{eq74} 3\rho H<0.
\end{equation}
From Eq. (\ref{eq74}), we observe that $H<0$, which gives the contradictory result and hence we conclude that cosmological constant
plays an important role on the evolution of the universe.
\section*{acknowledgements}
Author would like to thank the Islamic Azad University, Mahshahr branch
for providing facility and support where this work was carried out.

\end{document}